\documentclass[12pt,letterpaper]{article}
\usepackage[a4paper, total={7in, 10in}]{geometry}

\usepackage{graphicx}
\usepackage{helvet}
\usepackage{authblk}
\usepackage{hyperref}
\usepackage{amsmath} 
\usepackage{amssymb} 
\usepackage{orcidlink} 
\usepackage[super,comma,sort&compress]  
   {natbib}
\usepackage{multirow}%
\usepackage{rotating}
\usepackage{caption}
\usepackage{amsfonts}%
\usepackage{amsthm}%
\usepackage{mathrsfs}%
\usepackage{xcolor}%
\usepackage{textcomp}%
\usepackage{manyfoot}%
\usepackage{booktabs}%
\usepackage{algorithm}%
\usepackage{algorithmicx}%
\usepackage{algpseudocode}%
\usepackage{listings}%
\usepackage{longtable}
\usepackage{hyperref}  
\usepackage{url}
\usepackage{threeparttable}

\makeatletter
\renewcommand{\maketitle}{\bgroup\setlength{\parindent}{0pt}
\begin{flushleft}
  \textbf{\@title}
  
  \@author
\end{flushleft}\egroup}
\makeatother

\title{Learning collective multi-cellular dynamics from temporal scRNA-seq via a transformer-enhanced Neural SDE}

\date{}

\author[1,2,\orcidlink{0009-0004-1683-746X}]{Qi Jiang}
\author[3]{Lei Zhang}
\author[1,2]{Longquan Li}
\author[1,2,\orcidlink{0000-0002-3511-0512},*]{Lin Wan}

\affil[1]{State Key Laboratory of Mathematical Sciences, Academy of Mathematics and Systems Science, Chinese Academy of Sciences, No. 55 Zhongguancun East Road, 100190, Beijing, China}
\affil[2]{School of Mathematical Sciences, University of Chinese Academy of Sciences, 19A Yuquan Road, 100049, Beijing, China}
\affil[3]{Department of Control Science and Engineering, Tongji University, No. 4800 Cao'an Road, 201804, Shanghai, China}

\affil[*]{Correspondence: lwan@amss.ac.cn}

\begin{document}

\maketitle

\section*{SUMMARY}

Time-series single-cell RNA-sequencing (scRNA-seq) datasets offer unprecedented insights into the dynamics and heterogeneity of cellular systems. 
These systems exhibit multiscale collective behaviors driven by intricate intracellular gene regulatory networks and intercellular interactions of molecules. 
However, inferring interacting cell population dynamics from time-series scRNA-seq data remains a significant challenge, as cells are isolated and destroyed during sequencing. 
To address this, we introduce scIMF, a single-cell deep generative Interacting Mean Field model, designed to learn collective multi-cellular dynamics. 
Our approach leverages a transformer-enhanced stochastic differential equation network to simultaneously capture cell-intrinsic dynamics and intercellular interactions. 
Through extensive benchmarking on multiple scRNA-seq datasets, scIMF outperforms existing methods in reconstructing gene expression at held-out time points, demonstrating that modeling cell-cell communication enhances the accuracy of multicellular dynamics characterization. 
Additionally, our model provides biologically interpretable insights into cell-cell interactions during dynamic processes, offering a powerful tool for understanding complex cellular systems.

\section*{KEYWORDS}


time-series single-cell data, cell-cell interactions, attention  mechanism, neural SDE, transformer

\section*{INTRODUCTION}

Emerging time-series single-cell RNA sequencing (scRNA-seq) data offer unprecedented opportunities to study dynamic cellular processes, such as cell differentiation and tumorigenesis. 
However, leveraging time-series scRNA-seq data poses significant challenges due to the destructive nature of single-cell sequencing technologies \citep{yuan2017challenges}. 
During sequencing, cells are isolated and destroyed, resulting in the loss of cell-cell correspondences both within and across time points. 
Multicellular systems depend on the precise coordination of cellular activities, which are mediated by complex cell-cell interactions (CCIs) among diverse cell types \citep{Armingol2020DecipheringCI, Armingol2024CCC}. 
Consequently, modeling and inferring the collective dynamics of cell populations with nonlinear interactions from time-series scRNA-seq data remains a major computational challenge.

Extensive computational efforts have been made to link scRNA-seq snapshots taken at different time points. 
Methods leveraging the framework of optimal transport (OT) \citep{kantorovitch1958translocation} have been developed to learn optimal, cost-effective mappings between data distributions.   
Early attempts, such as Waddington-OT \citep{Schiebinger2019OptimalTransportAO} and a generative adversarial network by \citeauthor{Yang2018Scalable} \citep{Yang2018Scalable}, utilize unbalanced optimal transport frameworks to infer probabilistic couplings between cells at different time points.
 While effective for snapshot alignment, these methods are limited to static mappings.
Recent advances overcome this limitation by incorporating dynamic optimal transport (DOT) \citep{benamou2000computational}, enabling inference of continuous trajectories.  
For examples,
TrajectoryNet integrates DOT with continuous normalized flow to efficiently model evolving cell populations in continuous time \citep{Tong2020TrajectoryNetAD}; 
MIOFlow extends TrajectoryNet by combining stochastic dynamics with a latent manifold structure, learned via a geometric variational autoencoder \citep{Huguet2022ManifoldIO}.
Besides, PRESCIENT models cell dynamics as stochastic diffusion processes governed by an underlying potential energy landscape, akin to the Waddington landscape.
It employs neural networks to learn this potential-driven dynamics \citep{Yeo2021GenerativeMO};
PI-SDE further extends the framework of PRESCIENT by incorporating the principle of least action to derive a more interpretable cellular potential landscape \citep{PISDE}.
Alternatively, scNODE addresses the problem of distribution shifts between observed and unobserved time points by embedding neural ordinary differential equation (ODE) into a variational autoencoder, along with a dynamic regularization term \citep{scNODE}.

Despite these advances, current methods are limited in their ability to model cell-cell interactions or correlations of cells within and across snapshots. 
These approaches model the dynamics of each cell based on its own states, without explicitly incorporating information about surrounding or coexisting cells.
This limitation reduces their accuracy in predicting or generating collective cell population dynamics. 
Meanwhile, a growing body of evidence underscores the critical role of cell-cell interactions (CCIs) in regulating dynamic biological processes, such as development, immune responses, tissue repair, and disease progression \citep{Armingol2020DecipheringCI, Armingol2024CCC}. 
This raises a key question: How can intercellular relationships be effectively integrated into models of cell population dynamics to simultaneously capture temporal dynamics and cell-cell interactions?

To tackle this challenge, a handful of methods have been proposed to model the dynamics of cell populations that take into account cell-cell interactions.
For example, GraphFP models the dynamics of cell populations based on Wasserstein gradient flows, employing a nonlinear free energy functional  with a quadratic cell-cell interaction term \citep{jiang2022dynamic}.
Although biologically interpretable and computationally efficient, it  operates on discrete state space at aggregated cell type/cluster level, leading to a loss of finer resolution when inferring  the continuous cell state at single cell level.  DIISCO also works at the cluster level, inferring dynamic intercellular interactions using Gaussian processes guided by receptor-ligand pair priors \citep{DIISCO}. 
Smart and Zilman propose a statistical physics model based on Hopfield networks, capturing intra- and intercellular gene interactions \citep{SMART2023101247}. However, this approach relies on binarized gene expression and scales poorly with increasing cell/gene numbers.

To fill this gap, we propose scIMF, a single-cell deep generative Interacting Mean Field Model that jointly infers cellular dynamics and quantifies cell-cell interactions from time-series scRNA-seq data. 
Unlike traditional It\^o-type SDE approaches that typically adopt single-particle or agent-based paradigms, our method employs the McKean-Vlasov SDE (MV-SDE) \citep{Mckean1966ACO, MVE, dawson1983critical}, where each cell's state evolution depends on the population's empirical distribution, enabling explicit modeling of both temporal patterns and intercellular dependencies. 
Leveraging the Transformer's self-attention mechanism  \citep{Attention}, scIMF encodes the MV-SDE's interacting mean-field drift through a transformer  with cell-specific attention, efficiently capturing non-local relationships of cells. This integration of MV-SDE theory with modern deep learning allows scIMF to handle large-scale, high-dimensional data while holistically incorporating nonlocal cell-cell interactions, yielding improved prediction accuracy and more comprehensive representations of complex cellular dynamics.


We conducted comprehensive benchmarking of scIMF against state-of-the-art time-series scRNA-seq inference methods (PRESCIENT, MIOFlow, PI-SDE, and scNODE) across three datasets. 
Our results demonstrate that scIMF achieves superior performance in both key evaluation metrics: 
(1) accurately reconstructing gene expression populations at held-out time points, 
(2) inferring cellular velocities that better align with observed temporal transition patterns. 
These findings provide strong empirical evidence that explicitly modeling cell-cell interactions leads to more accurate representations of multicellular system dynamics compared to methods that model the dynamics of each cell based on its own states. 
Besides, the estimated cell-wise attention scores is asymmetric, reflecting a lack of reciprocity in cell-cell interactions and communications in the out-of-equilibrium cell systems.

\section*{RESULTS}

\subsection*{Overview of scIMF}
\begin{figure*}[!t]%
\centering
\includegraphics[width=0.90\linewidth]{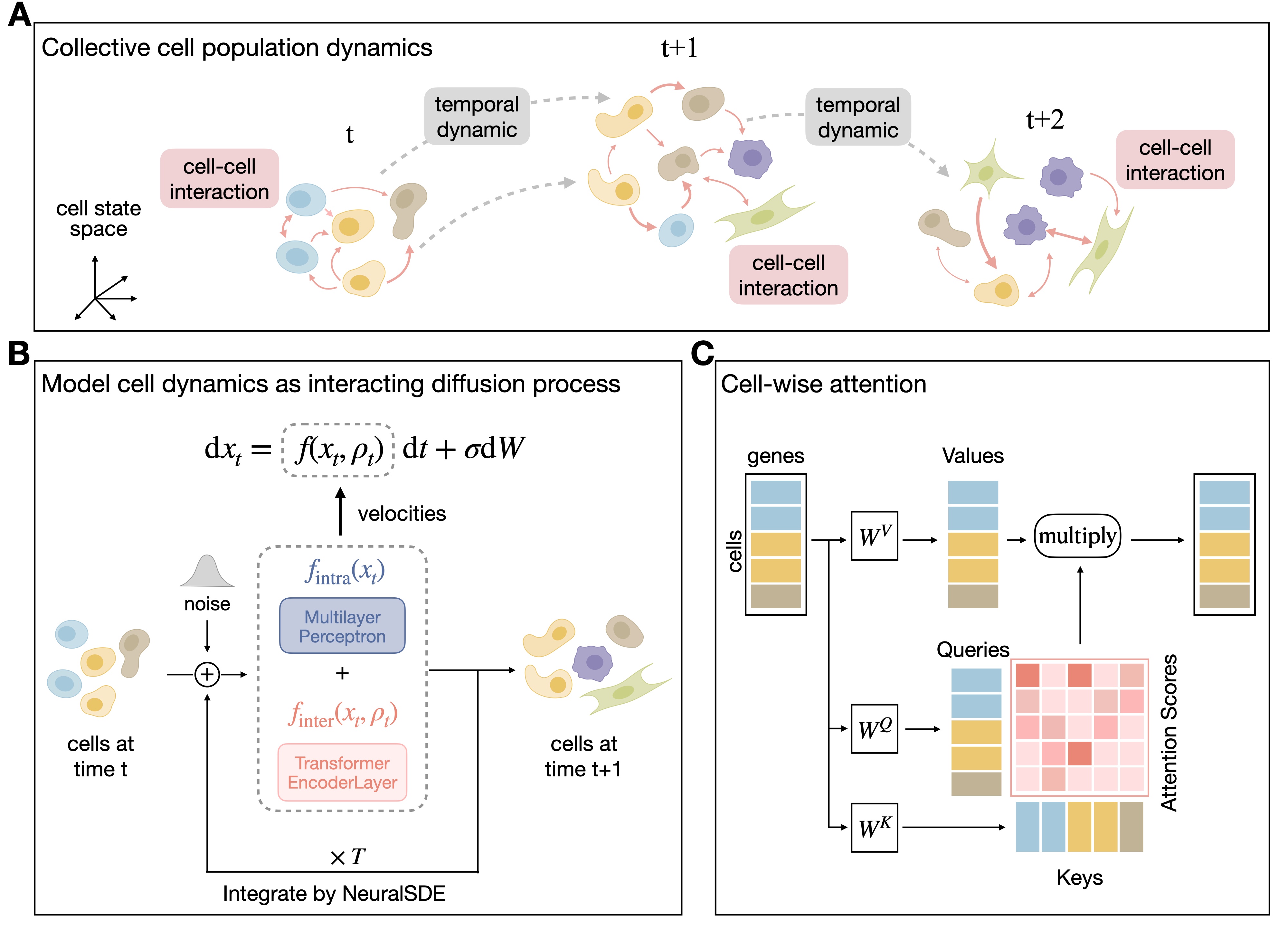}
\caption{ Overview of scIMF. (A) For time-series data describing the collective cell population dynamics, the destructive nature of scRNA-seq techniques results in a lack of cell-cell correspondence both within time point and across time points. (B) scIMF models complex multicellular dynamics as interacting diffusion processes, described by the MV-SDE, and solves the MV-SDE using the NeuralSDE framework, where the drift coefficient is approximated through two neural networks. (C) scIMF encodes the cell-cell interactions through a cell-wise attention architecture. 
}
\label{fig:overview}
\end{figure*}

scIMF is a deep generative network that learns the collective dynamics of cell populations from time-series scRNA-seq data (Fig.~\ref{fig:overview}).
To achieve this, it models complex multicellular dynamics as interacting diffusion processes, described by the MV-SDE. 
MV-SDE gives rise to nonlinear and nonlocal Kolmogorov forward equations, which are particularly suited to describe interacting cellular systems \citep{MVE}. 
Moreover, we utilize the self-attention architecture of Transformer, which are well-suited for capturing non-local relationships between tokens (e.g.,  cells in our study). 
By treating gene expression vector of each cell as tokens, we implement a cell-wise attention mechanism to captures cell-cell interactions. 
Finally, the entire model is embedded within the DOT framework and efficiently solved by the Neural SDE \citep{li2020scalable,kidger2021neuralsde}, with a transformer encoder network approximating the interacting mean-field drift term of MV-SDE.
Notably, scIMF is both model- and data-driven, requiring no prior biological knowledge (e.g., predefined receptor-ligand pairs). Given time-series scRNA-seq data, the model can reconstruct developmental trajectories over time while simultaneously quantifying time-varying cell-cell interactions.

To comprehensively evaluate our method, 
we benchmark scIMF against four state-of-the-art methods, 
PRESCIENT, MIOFlow, PI-SDE, and scNODE, 
using three time-series scRNA-seq datasets: 
(1) zebrafish embryogenesis \citep{zebrafishData} (denoted as ZB data), 
(2) reprogramming of mouse embryonic fibroblasts to induced pluripotent stem cells \citep{wotData} (denoted as MEF data),
and (3) pancreatic $\beta$-cell differentiation \citep{veresData} (denoted as Panc data).
The ZB and MEF datasets are collected from in vivo experiments and were included in the study of scNODE \citep{scNODE}, 
while the Panc dataset is collected from in vitro experiment and has been used in prior work including PRESCIENT and PI-SDE \citep{Yeo2021GenerativeMO, PISDE}.

\subsection*{scIMF accurately predicts gene expression profiles in ZB and MEF datasets}

First, we assess the predictive ability of scIMF on ZB and MEF datasets from in vivo experiments, and compare it with four state-of-the-art baselines: PRESCIENT, MIOFlow, PI-SDE, and scNODE.  
Specifically, we evaluate how well each method predicts gene expression profiles at unseen time points.

To this end, we conduct held-out tasks on the ZB and MEF datasets. During training, data from multiple time points are excluded. After training, each model is evaluated by comparing its predictions on the excluded (unseen) time points with the true expression profiles using $L_1$- and $L_2$-Wasserstein distances ($\mathcal{W}_1$ and $\mathcal{W}_2$), where smaller values indicate better performance.
Following the protocol in scNODE \citep{scNODE}, we use three types of held-out tasks:
(i) easy tasks, where data at middle time points are removed to evaluate interpolation; 
(ii) medium tasks, where data at the last few time points are removed to evaluate extrapolation;
and (iii) hard tasks, which combines both schemes of easy and medium tasks to assess the model's performance under simultaneous interpolation and extrapolation challenges.
Details of the held-out time points for each task and dataset are summarized in Table~\ref{tab:held-out tasks}.

\begin{table}[h]
\centering
\small
\caption{Held-out time points for different tasks across datasets.}
\label{tab:held-out tasks}
\begin{tabular}{|l|c|c|c|c|} 
 \hline
 \textbf{Dataset} & \textbf{\#Timepoints} & \textbf{Easy Tasks} & \textbf{Medium Tasks} & \textbf{Hard Tasks} \\ [1ex] 
 \hline
 ZB  & 12  & 4, 6, 8 & 10, 11 & 2, 4, 6, 8, 10, 11 \\ [1ex] 
 \hline
 MEF & 19  & 5, 10, 15 & 16, 17, 18 & 5, 7, 9, 11, 15, 16, 17, 18 \\ [1ex] 
 \hline
\end{tabular}
\begin{tablenotes}
\footnotesize
\item Time points are indexed starting from 0.
\end{tablenotes}
\end{table}

As shown in Table~\ref{tab: Held-out ZB}, scIMF consistently achieves the lowest Wasserstein distances across all tasks on ZB data. On the easy task, scIMF reaches $\mathcal{W}_1 = 10.81, \mathcal{W}_2 = 12.25$, outperforming all baselines. The improvement becomes more pronounced on medium and hard tasks, where extrapolation is included. For example, on the hard task, scIMF reduces the error to $\mathcal{W}_1 = 12.92, \mathcal{W}_2 = 15.77$, which corresponds to a 17.2\% and 13.5\% reduction in Wasserstein distances compared to the closest competitor, scNODE ($\mathcal{W}_1 = 15.66, \mathcal{W}_2 = 18.23$).

\begin{table}[h]
\centering
\small
\begin{threeparttable}
\caption{Held-out performance for different tasks on ZB data.}
\label{tab: Held-out ZB}
\begin{tabular}{|l|c|c|c|c|c|c|}
\hline
\textbf{Model} & \multicolumn{2}{c|}{\textbf{Easy Task}} & \multicolumn{2}{c|}{\textbf{Medium Task}} & \multicolumn{2}{c|}{\textbf{Hard Task}} \\
\cline{2-7}
& $\mathcal{W}_1$ & $\mathcal{W}_2$ & $\mathcal{W}_1$ & $\mathcal{W}_2$ & $\mathcal{W}_1$ & $\mathcal{W}_2$ \\ [1ex]
\hline
PRESCIENT  & 12.38 ± 0.05 & 14.72 ± 0.04 & 29.11 ± 0.50 & 36.92 ± 1.15 & 16.41 ± 0.40 & 21.21 ± 1.13 \\ [1ex]
\hline
MIOFlow    & 12.17 ± 0.06 & 14.61 ± 0.07 & 29.92 ± 4.27 & 34.73 ± 5.75 & 16.68 ± 1.21 & 19.48 ± 0.64 \\ [1ex]
\hline
PI-SDE     & 11.39 ± 0.12 & 14.38 ± 0.18 & 27.09 ± 1.61 & 35.79 ± 3.41 & 15.79 ± 1.24 & 21.17 ± 3.00 \\ [1ex]
\hline
scNODE     & 11.71 ± 0.02 & 14.60 ± 0.001 & 22.89 ± 0.60 & 26.57 ± 0.62 & 15.66 ± 0.05 & 18.23 ± 0.18 \\ [1ex]
\hline
\textbf{scIMF} & \textbf{10.81} ± \textbf{0.01} & \textbf{12.25} ± \textbf{0.03} & \textbf{17.22} ± \textbf{0.05} & \textbf{20.13} ± \textbf{0.26} & \textbf{12.92} ± \textbf{0.03} & \textbf{15.77} ± \textbf{0.36} \\ [1ex]
\hline
\end{tabular}
\begin{tablenotes}
\footnotesize
\item Held-out tasks were evaluated using $L_1$- and $L_2$-Wasserstein distances ($\mathcal{W}_1$ and $\mathcal{W}_2$) between true and predicted gene expression profiles. Results are averaged over 5 seeds.
\item Bold values indicate the best performance across all models for each task.
\end{tablenotes}
\end{threeparttable}
\end{table}

On MEF data (Table~\ref{tab: Held-out MEF}), scIMF again achieves the best average performance across all three tasks. Particularly, on the medium task, which involves extrapolation over several late time points, scIMF outperforms others with $\mathcal{W}_1 = 21.14, \mathcal{W}_2 = 29.88$, compared to scNODE ($\mathcal{W}_1 = 26.51, \mathcal{W}_2 = 33.26$) and PI-SDE ($\mathcal{W}_1 = 29.05, \mathcal{W}_2 = 36.05$). On the hard task, scIMF achieves the lowest $\mathcal{W}_1 = 18.90$. While its $\mathcal{W}_2 = 27.73$ is slightly higher than that of scNODE ($25.98$), it remains comparable to PI-SDE ($27.11$) and PRESCIENT ($27.92$).

\begin{table}[h]
\centering
\small
\begin{threeparttable}
\caption{Held-out performance for different tasks on MEF data.}
\label{tab: Held-out MEF}
\begin{tabular}{|l|c|c|c|c|c|c|}
\hline
\textbf{Model} & \multicolumn{2}{c|}{\textbf{Easy Task}} & \multicolumn{2}{c|}{\textbf{Medium Task}} & \multicolumn{2}{c|}{\textbf{Hard Task}} \\
\cline{2-7}
& $\mathcal{W}_1$ & $\mathcal{W}_2$ & $\mathcal{W}_1$ & $\mathcal{W}_2$ & $\mathcal{W}_1$ & $\mathcal{W}_2$ \\ [1ex]
\hline
PRESCIENT  & 13.02 ± 0.07 & 14.51 ± 0.07 & 30.28 ± 0.42 & 35.76 ± 0.38 & 23.14 ± 0.11 & 27.92 ± 0.31 \\ [1ex]
\hline
MIOFlow    & 15.06 ± 0.36 & 16.43 ± 0.42 & 28.82 ± 1.27 & 35.76 ± 1.62 & 25.06 ± 3.11 & 29.70 ± 2.16 \\ [1ex]
\hline
PI-SDE     & 12.20 ± 0.07 & 13.80 ± 0.09 & 29.05 ± 0.12 & 36.05 ± 0.30 & 21.92 ± 0.08 & 27.11 ± 0.02 \\ [1ex]
\hline
scNODE     & 13.96 ± 0.03 & 15.45 ± 0.06 & 26.51 ± 0.10 & 33.26 ± 0.72 & 21.02 ± 0.07 & \textbf{25.98} ± 0.26 \\ [1ex]
\hline
\textbf{scIMF} & \textbf{11.83} ± \textbf{0.01} & \textbf{13.79} ± \textbf{0.02} & \textbf{21.14} ± \textbf{0.14} & \textbf{29.88} ± \textbf{0.89} & \textbf{18.90} ± \textbf{0.16} & 27.73 ± 0.88 \\ [1ex]
\hline
\end{tabular}
\begin{tablenotes}
\footnotesize
\item Held-out tasks were evaluated using $L_1$- and $L_2$-Wasserstein distances ($\mathcal{W}_1$ and $\mathcal{W}_2$) between true and predicted gene expression profiles. Results are averaged over 5 seeds.
\item Bold values indicate the best performance across all models for each task.
\end{tablenotes}
\end{threeparttable}
\end{table}

\subsection*{scIMF recapitulates complex developmental branching and cellular velocities in ZB and MEF datasets}

\begin{figure*}[htbp]%
\centering
\includegraphics[width=0.95\linewidth]{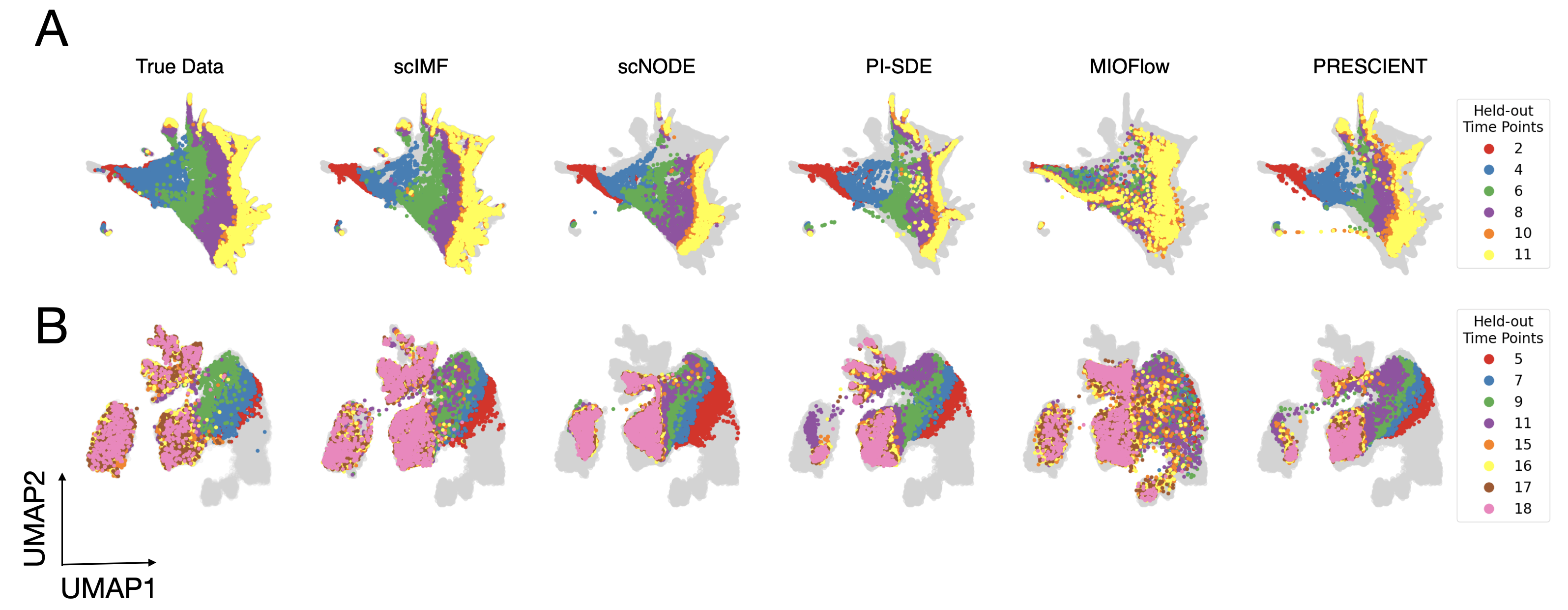}
\caption{UMAPs of true and predicted gene expressions for hard held-out task on ZB data (A) and MEF data (B), with gray points representing observed data.}
\label{fig:UMAP-hard}
\end{figure*}

We further visualize the true gene expressions and model predictions for the hard held-out tasks across datasets using Uniform Manifold Approximation and Projection (UMAP) \citep{mcinnes2018umap}.
As shown in Fig.~\ref{fig:UMAP-hard}, the predicted gene expression profiles generated by scIMF align with the ground truth well.
This suggests that the incorporation of cell-cell interactions enables scIMF to capture complex multicellular dynamics.
Notably, for ZB data, characterized by its multi-branching developmental processes during zebrafish embryogenesis, scIMF accurately predicts fine-grained cell fates at later stages.
In contrast, baseline methods, such as scNODE and PI-SDE, fail to capture the full spectrum of cell fate outcomes, often inferring only a limited number of terminal states. By modeling the complex branching structure, scIMF provides a more detailed and precise reconstruction of cellular fates.

\begin{figure*}[htbp]
\centering
\includegraphics[width=0.90\linewidth]{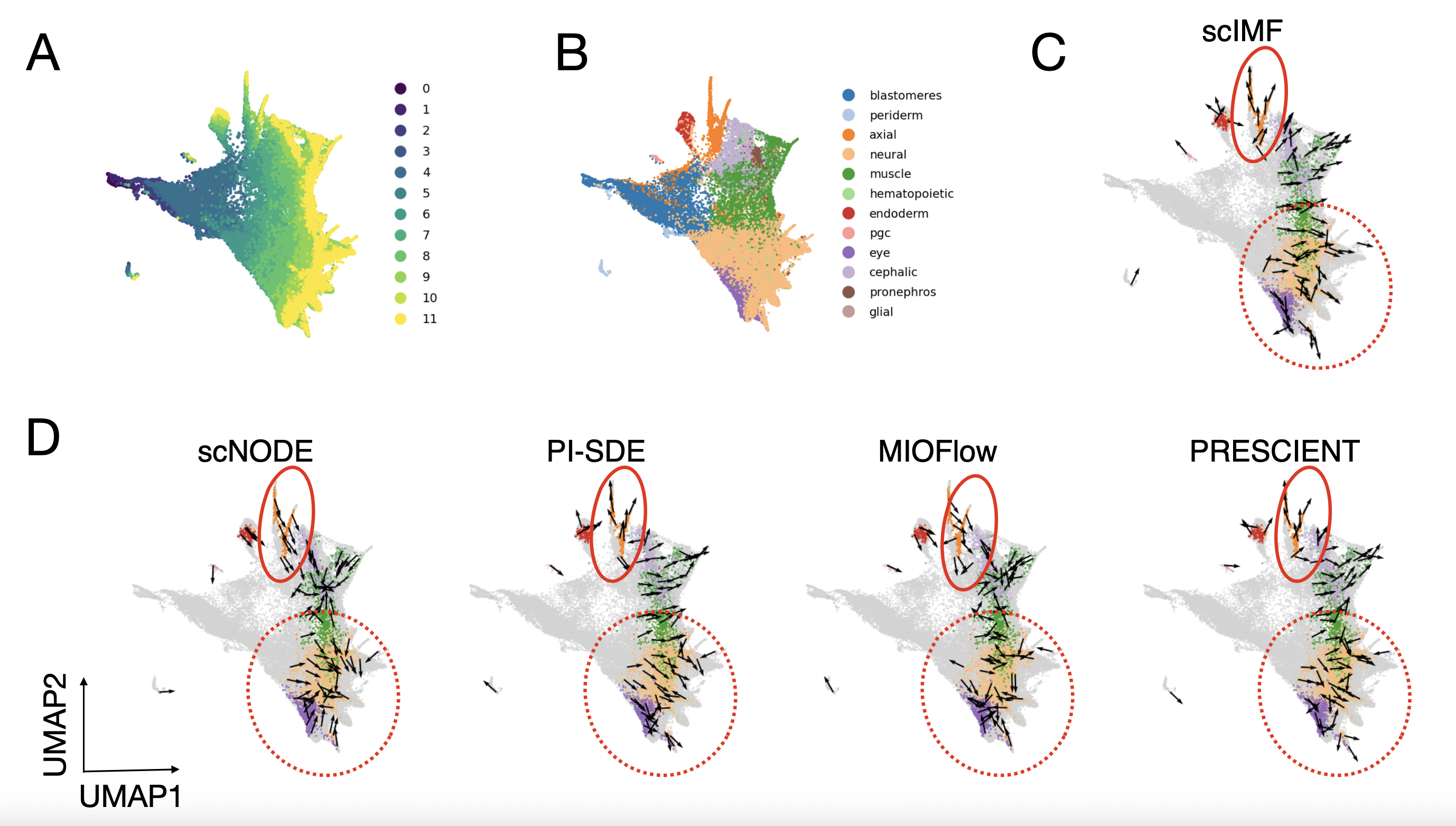}
\caption{scIMF's performance on ZB data. (A)UMAP of zebrafish embryogenesis cells, colored by developmental time points. (B)UMAP of zebrafish embryogenesis cells, colored by cell types. (C,D)Velocities for a random sample of observed cells at $t=8$ inferred by scIMF and other baseline methods.}
\label{fig:zebrafish}
\end{figure*}

To further illustrate the superiority of scIMF, we focus on the hard held-out task on ZB data, which comprises 12 developmental stages and 12 cell types (Fig.~\ref{fig:zebrafish}A,B). 
We evaluate how well the cellular velocities inferred by different methods align with the expected temporal progression. Specifically, we visualize the cellular velocities predicted by scIMF and other baseline methods at time point $t=8$ (Fig.~\ref{fig:zebrafish}C,D).
Notably, scIMF accurately drives cells toward multiple terminal fates, particularly for neural and eye cells (region within the red dashed line) as well as axial cells (region within the red solid line).
In contrast, while PI-SDE and PRESCIENT perform well in the axial cell region, their predictions in the neural and eye regions exhibit substantial deviations from the observed temporal pattern in Fig.~\ref{fig:zebrafish}A.
Furthermore, scNODE and MIOFlow underperform in both regions, with their predicted velocities failing to match the observed developmental directions


\subsection*{scIMF reveals non-reciprocal cell-cell interactions in ZB dataset from in vivo experiment}

\begin{figure*}[htbp]%
\centering
\includegraphics[width=0.85\linewidth]{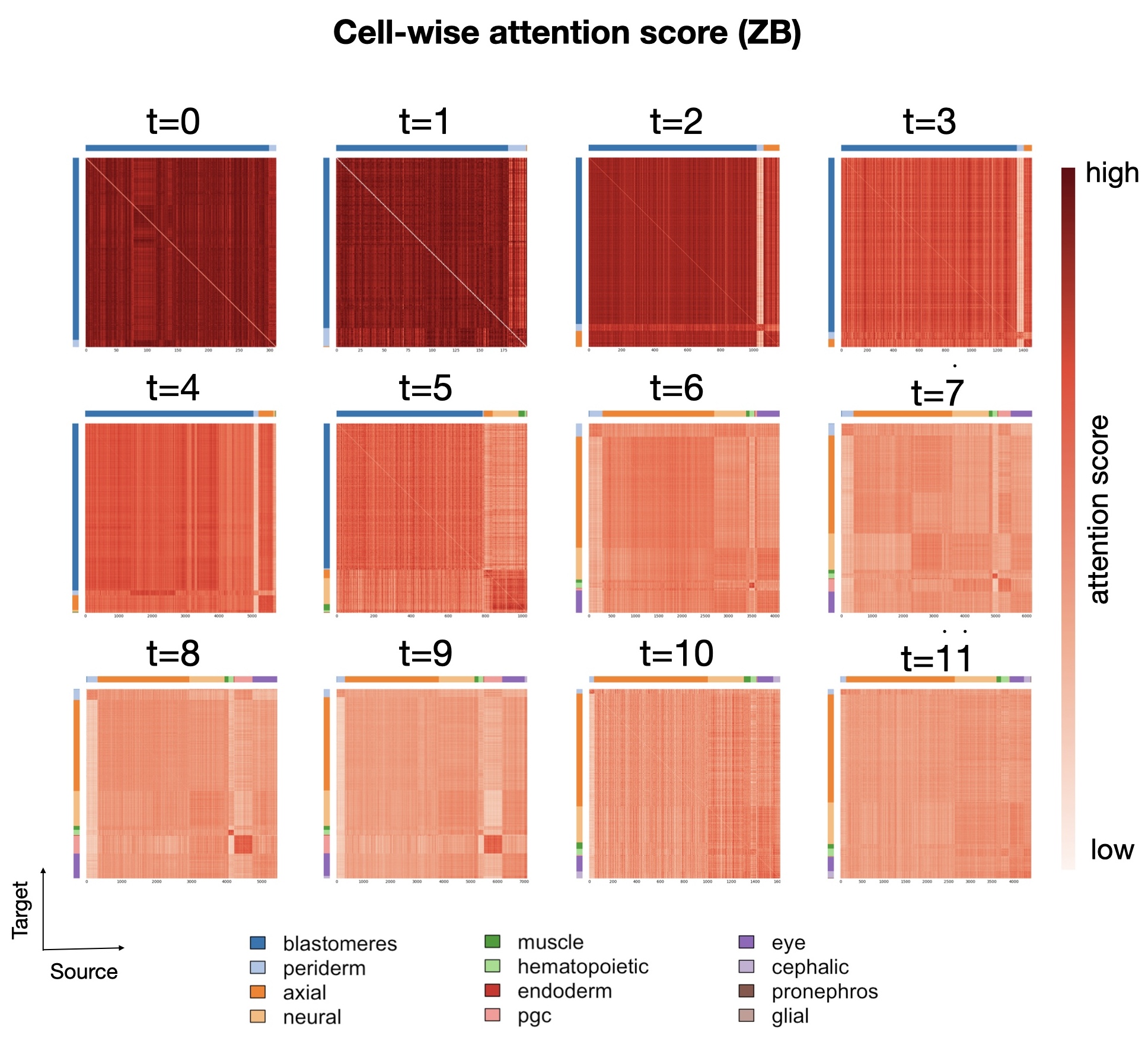}
\caption{Cell-wise attention scores on ZB data.}
\label{fig:Attention-ZB}
\end{figure*}

A key feature of scGenIMF is its use of a cell-wise attention mechanism, which captures intercellular relationships.
The attention score at position $(i, j)$ quantifies the influence of cell $j$ on cell $i$ within the context of collective dynamics. 
When the score approaches zero, it suggests that cell $j$ has negligible impact on cell $i$'s dynamics, indicating minimal interaction or communication between the two cells.
To explore this further, we analyze the estimated cell-wise attention scores over time (Fig.~\ref{fig:Attention-ZB}). 
A notable observation is the progressive emergence of clearer block structures in the attention matrices as development proceeds. 
Each block reflects the interaction pattern from one specific cell type to another.
These patterns highlight how different cell types communicate with each other across time. 
Importantly, these attention matrices are non-symmetric, allowing scIMF to capture non-reciprocal relationships between cells. This asymmetry reveals valuable insights into the directionality of cell-cell communication and interaction dynamics.

In the early stages of development ($t=0, 1,2, 3$),  attention scores within interactions involving the same cell type often exhibit high heterogeneity.
For instance, blastomeres cells show diverse sub-population behaviors, suggesting complex and variable interaction patterns. 
 As development advances to later stages ($t=7, 8,9,10$), these interactions within cell-type become more homogeneous.  This shift is particularly evident in neural cells, indicating their transition into more functionally specialized populations.
Meanwhile, interactions between different cell types are particularly active during the early stages of zebrafish embryonic development ($t=0$ to $t=5$),  and significantly diminish in later stages.
This change suggests that communication between different cell types decreases as development progresses. 
Overall, these results reveal a progressive transition from global inter-lineage communication in early development to localized, lineage-specific attention patterns, reflecting the underlying biological progression from pluripotency to functional specialization.

To further investigate how attention is distributed among cell types, we compute the cell type-wise attention scores by averaging cell-wise attention scores according to annotated cell type labels (Fig.~\ref{fig:Attention-ZB}). 
In each diagram,  the arrows denote the directional influence from the source cell type to the target cell type. The color of each arrow corresponds to the color of the source cell type, and the arrow width is proportional to the magnitude of the mean attention score. 

During the early developmental stages ($t=0$ to $t=3$), attention patterns are characterized by widespread and relatively uniform interactions among a few early lineages, namely blastomeres, periderm, and axial. These lineages exhibit strong bidirectional attention scores, indicating a phase of broad regulatory potential and high developmental plasticity.
Starting from $t=4$, newly emerging cell types including neural, muscle, hematopoietic, and endoderm, begin to appear. Concurrently, the interaction patterns start to exhibit more lineage-specific structures. 
For example, at $t=4$ and $t=5$, neural cells primarily act as recipients, receiving signals mainly from muscle and endoderm. In terms of interaction strength, endoderm, hematopoietic, and muscle cells exhibit strong outgoing interactions, indicating their active roles as source cell types during this stage.
Meanwhile, the attention scores emanating from blastomeres and periderm cells gradually decline.
From $t = 10$ to $t = 11$, the interactions among cell types start to become homogeneous again, with bidirectional interactions. This corresponds to the uniformity of the matrices in cell-wise attention at $t = 10, 11$.

\begin{figure*}[htbp]%
\centering
\includegraphics[width=0.85\linewidth]{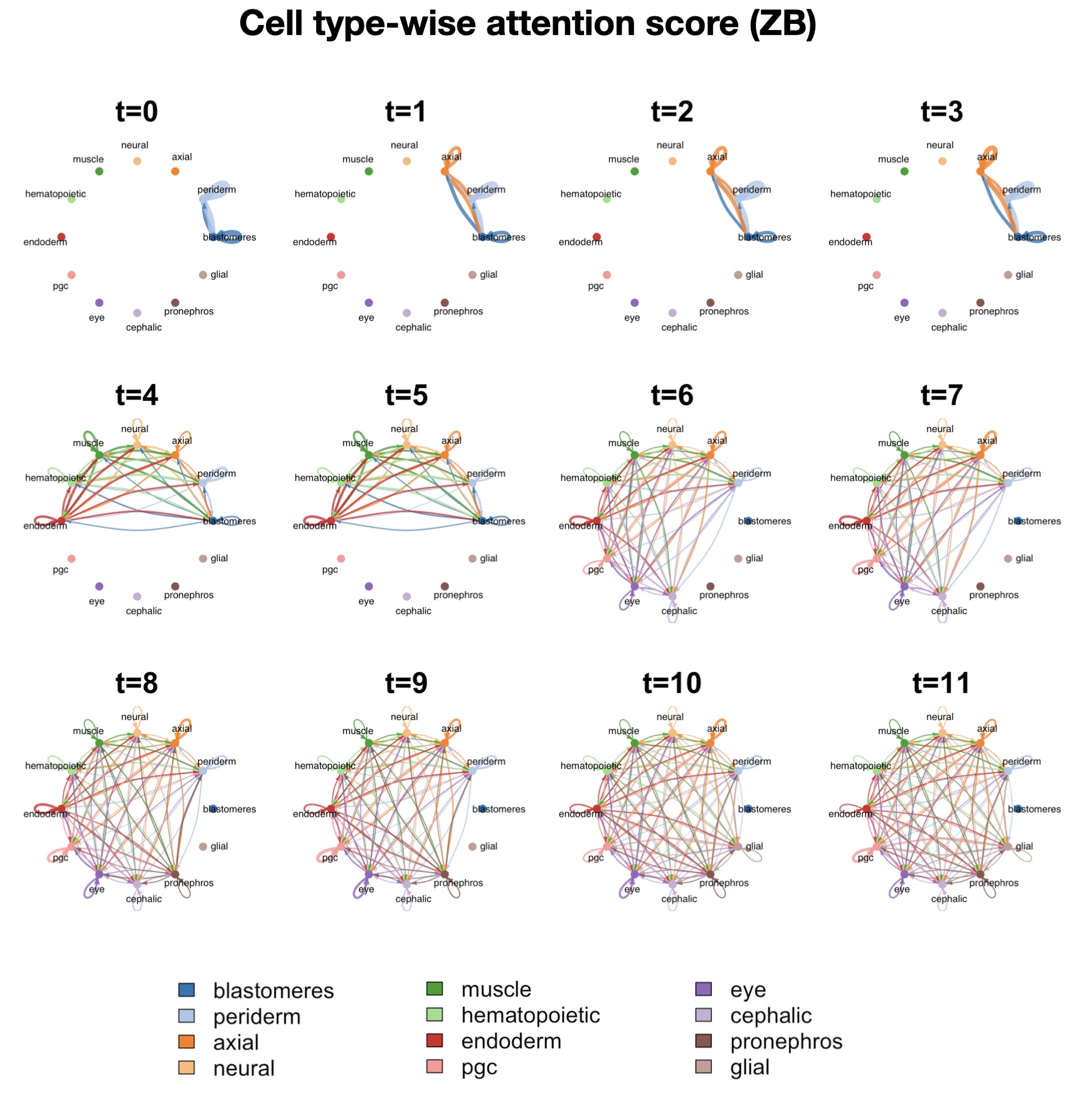}
\caption{Cell type-wise attention scores on ZB data. The arrow denotes the directional influence from the source cell type to the target cell type. The color of arrow corresponds to the color of the source cell type. The arrow width is proportional to the attention scores.}
\label{fig:Chord-ZB}
\end{figure*}


\subsection*{scIMF achieve high accuracy in Panc dataset and reveal symmetry interaction pattern}


To assess the applicability of scIMF in a different biological context of in vitro experiment, we evaluate its performance on a time-series single-cell dataset of in vitro $\beta$-cell differentiation (Panc data).
We conduct held-one-timepoint-out prediction tasksby sequentially holding out each time point beyond the first and second time points, and compare the predictive performance of scIMF with other baseline methods (Table~\ref{tab:Held-out Panc1}, \ref{tab:Held-out Panc2}).
Overall, PI-SDE achieves the highest accuracy, and scGeneIMF achieves comparable reuslts as PI-SDE, with the second highest accuracy.

\begin{table}[h]
\centering
\small
\begin{threeparttable}
\caption{Held-out performance for Panc data (time points $t_2$ to $t_4$).}
\label{tab:Held-out Panc1}
\begin{tabular}{|l|c|c|c|c|c|c|}
\hline
\textbf{Model} & \multicolumn{2}{c|}{\textbf{Held-out $t_2$}} & \multicolumn{2}{c|}{\textbf{Held-out $t_3$}} & \multicolumn{2}{c|}{\textbf{Held-out $t_4$}} \\
\cline{2-7}
& $\mathcal{W}_1$ & $\mathcal{W}_2$ & $\mathcal{W}_1$ & $\mathcal{W}_2$ & $\mathcal{W}_1$ & $\mathcal{W}_2$ \\ [1ex]
\hline
PRESCIENT  & 10.14 $\pm$ 0.01 & 10.97 $\pm$ 0.01 & 8.88 $\pm$ 0.01 & 9.35 $\pm$ 0.02 & 9.18 $\pm$ 0.01 & 9.70 $\pm$ 0.01 \\ [1ex]
\hline
MIOFlow    & 12.26 $\pm$ 0.01 & 13.48 $\pm$ 0.01 & 10.95 $\pm$ 0.04 & 11.71 $\pm$ 0.04 & 11.77 $\pm$ 0.04 & 12.58 $\pm$ 0.07 \\ [1ex]
\hline
PI-SDE     & \textbf{9.34} $\pm$ \textbf{0.05} & \textbf{10.02} $\pm$ \textbf{0.08} & \textbf{8.58} $\pm$ \textbf{0.01} & \textbf{9.00} $\pm$ \textbf{0.01} & \textbf{8.48} $\pm$ \textbf{0.02} & \textbf{8.94} $\pm$ \textbf{0.01} \\ [1ex]
\hline
scNODE     & 10.65 $\pm$ 0.06 & 11.48 $\pm$ 0.08 & 9.84 $\pm$ 0.01 & 10.35 $\pm$ 0.01 & 10.36 $\pm$ 0.04 & 11.05 $\pm$ 0.04 \\ [1ex]
\hline
\textbf{scIMF} & 11.68 $\pm$ 0.01 & 12.52 $\pm$ 0.01 & 9.41 $\pm$ 0.01 & 9.90 $\pm$ 0.01 & 8.94 $\pm$ 0.01 & 9.26 $\pm$ 0.01  \\ [1ex]
\hline
\end{tabular}
\begin{tablenotes}
\footnotesize
\item Bold values indicate the best performance.
\end{tablenotes}
\end{threeparttable}
\end{table}

\begin{table}[h]
\centering
\small
\begin{threeparttable}
\caption{Held-out performance for Panc data (time points $t_5$ to $t_7$).}
\label{tab:Held-out Panc2}
\begin{tabular}{|l|c|c|c|c|c|c|}
\hline
\textbf{Model} & \multicolumn{2}{c|}{\textbf{Held-out $t_5$}} & \multicolumn{2}{c|}{\textbf{Held-out $t_6$}} & \multicolumn{2}{c|}{\textbf{Held-out $t_7$}} \\
\cline{2-7}
& $\mathcal{W}_1$ & $\mathcal{W}_2$ & $\mathcal{W}_1$ & $\mathcal{W}_2$ & $\mathcal{W}_1$ & $\mathcal{W}_2$ \\ [1ex]
\hline
PRESCIENT  & 9.41 $\pm$ 0.02 & 9.92 $\pm$ 0.03 & 9.50 $\pm$ 0.01 & 10.01 $\pm$ 0.01 & 10.31 $\pm$ 0.01 & 10.86 $\pm$ 0.01 \\ [1ex]
\hline
MIOFlow    & 12.06 $\pm$ 0.03 & 12.85 $\pm$ 0.07 & 11.77 $\pm$ 0.01 & 12.51 $\pm$ 0.01 & 12.80 $\pm$ 0.07 & 13.79 $\pm$ 0.09\\ [1ex]
\hline
PI-SDE     & \textbf{8.64} $\pm$ \textbf{0.02} & \textbf{9.09} $\pm$ \textbf{0.03} & \textbf{8.78} $\pm$ \textbf{0.01} & \textbf{9.23} $\pm$ \textbf{0.01} & \textbf{9.31} $\pm$ \textbf{0.01} & \textbf{9.78} $\pm$ \textbf{0.01} \\ [1ex]
\hline
scNODE     & 10.72 $\pm$ 0.03 & 11.38 $\pm$ 0.02 & 10.62 $\pm$ 0.02 & 11.32 $\pm$ 0.02 & 11.17 $\pm$ 0.01 & 11.93 $\pm$ 0.01  \\ [1ex]
\hline
\textbf{scIMF} & 9.02 $\pm$ 0.01 & 9.58 $\pm$ 0.01 & 8.96 $\pm$ 0.01 & 9.34 $\pm$ 0.01 & 9.70 $\pm$ 0.01 & 10.19 $\pm$ 0.01  \\ [1ex]
\hline
\end{tabular}
\begin{tablenotes}
\footnotesize
\item Bold values indicate the best performance.
\end{tablenotes}
\end{threeparttable}
\end{table}

\begin{figure*}[htbp]%
\centering
\includegraphics[width=0.85\linewidth]{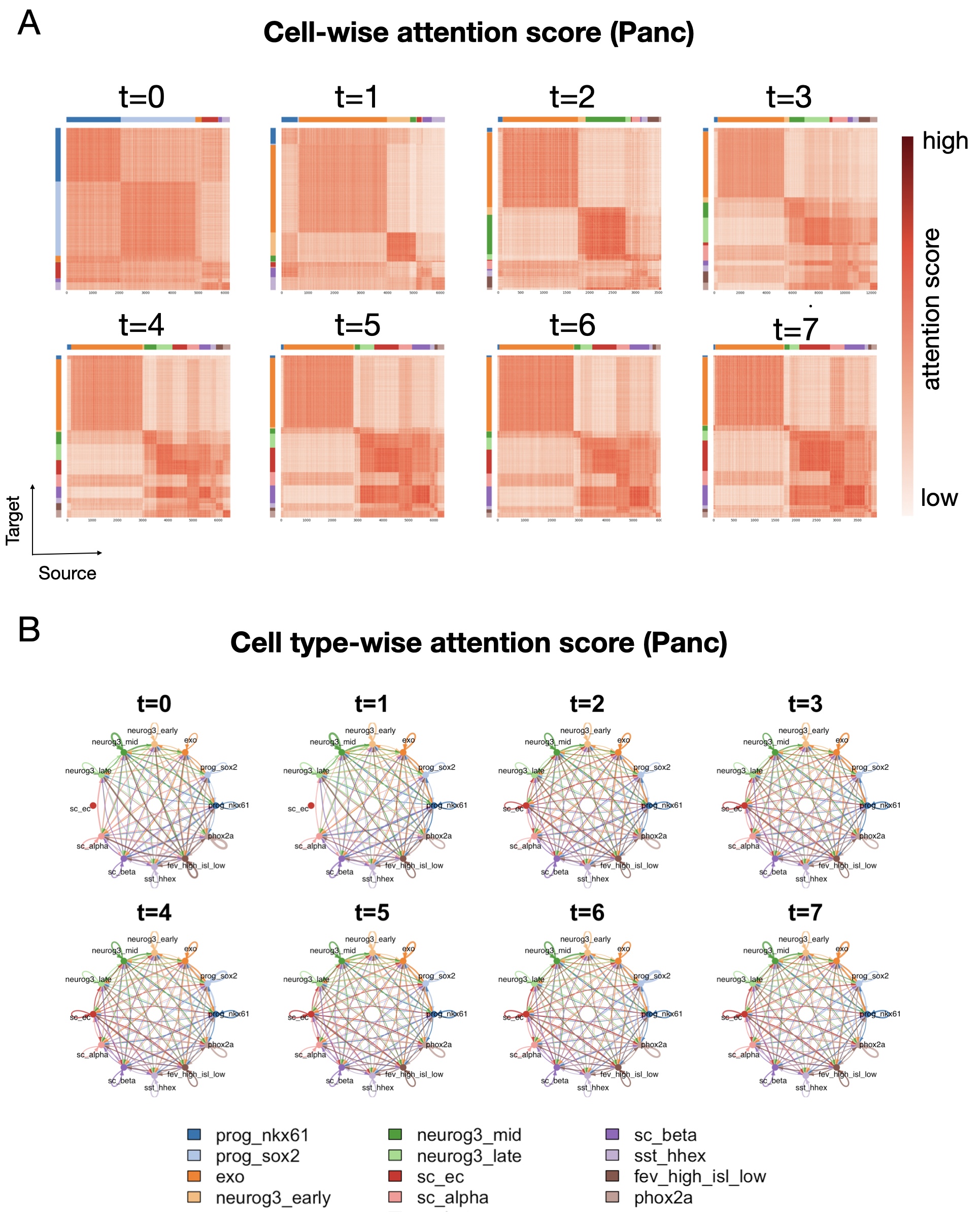}
\caption{The attention scores estimated by scIMF.  (A) Cell-wise attention scores on Panc  data.  (B) Cell type-wise attention score on Panc data. The arrow denotes the directional influence from the source cell type to the target cell type. The color of arrow corresponds to the color of the source cell type. The arrow width is proportional to the attention scores.}
\label{fig:Attention-Panc}
\end{figure*}

To better understand this result, we visualize the attention scores estimated by scIMF (Fig.~\ref{fig:Attention-Panc}). 
As shown in Fig.~\ref{fig:Attention-Panc}A, compared to those observed in zebrafish embryogenesis, the cell-wise attention matrices across all time points exhibit a relatively symmetric structure, suggesting that cell-cell interactions in this in vitro system are less intricate and more homogeneous.
Similarly, the cell type-wise attention scores in Fig.~\ref{fig:Attention-Panc}B show highly consistent interaction patterns from $t=2$ onward.
Slight changes are observed, mostly driven by the gradual emergence of specific populations such as \textit{neurog3\_mid} and \textit{neurog3\_late}.
From $t=4$ to $t=7$, the global interaction structures show minimal variation, which is consistent with the relatively stationary attention matrix patterns in Fig.~\ref{fig:Attention-Panc}A at the same stages.

Together, these observations indicate that the underlying dynamics of in vitro $\beta$-cell differentiation in this dataset are close to equilibrium or quasi-equilibrium, 
attributing to the reciprocal cell-cell interactions \citep{fruchart2021non}.
Under such conditions, PI-SDE appears sufficient to capture the overall system behavior, as reflected in its competitive performance. 
While scIMF is highly capable of learning the out of equilibrium systems with nonlinear dynamics and cell-cell interactions,
its advantages may be less pronounced in dynamic processes characterized by weak or stable interaction patterns. 


\section*{DISCUSSION}

Modelling collective cell dynamics has long been a central challenge in systems biology.
The destructive nature of scRNA-seq techniques leaves cell-cell correspondences unknown, both within a single snapshot (at the same time point) and between snapshots across different time points. This significantly complicates the reconstruction of multicellular dynamics over time.
To address these challenges, we propose scIMF, a deep generative interacting mean field model, that jointly infers cellular dynamics and quantifies cell-cell interactions from time-series scRNA-seq data. 
By implementing MV-SDEs where drift terms depend on the empirical cell-state distribution, we move beyond single-particle 
It$\hat{o}$-SDE approaches and explicitly model how individual fate decisions emerge from multicellular dynamics, capturing the collective behavior of the entire system that single-agent models typically fail to describe.
Since the coefficients of MV-SDE depends on the empirical distribution of the entire population, scIMF models how cells make fate decisions influenced by intercellular interactions, facilitating a more accurate reconstruction of dynamic processes.

Another key innovation of scIMF is its cell-wise attention mechanism for capturing cell-cell interactions. Unlike conventional neural networks, where nonlinearity arises solely from activation functions, the attention mechanism captures higher-order nonlinearities, distinguishing scIMF from classical It$\hat{o}$-type SDE-based methods. 
By combining the NeuralSDE framework with a transformer encoder layer to approximate the interacting mean field drift term, scIMF efficiently models not only the temporal evolution of cell populations but also the intercellular relationships. This integration results in an effective and interpretable tool for modeling collective cell dynamics and dynamic interactions at the individual cell level, offering a comprehensive representation of cellular systems.

Nonetheless, several aspects remain challenging and warrant further investigation. 
Our model employs a self-attention mechanism that effectively treats the cellular relationship graph as a fully connected graph, without explicitly incorporating its intrinsic topology.
The emergence of advanced single-cell technologies, such as spatio-temporal transcriptomics and lineage tracing via DNA barcoding, provides complementary insights into cellular dynamics. 
These techniques provide complementary information: spatio-temporal transcriptomics captures gene expression within spatial and temporal contexts, while lineage tracing uncovers clonal relationships and cell fate trajectories. 
A key challenge moving forward is how to leverage these data types to impose more structured constraints on the cell-cell relationship graph, potentially enhancing the biological fidelity of learned interactions. 
Developing methods that integrate such prior knowledge while maintaining the flexibility of attention-based architectures remains an exciting and open research direction.


\newpage


\section*{RESOURCE AVAILABILITY}

\subsection*{Lead contact}

Requests for further information and resources should be directed to and will be fulfilled by the lead contact, Lin Wan (lwan@amss.ac.cn).

\subsection*{Materials availability}

This study did not generate new unique reagents

\subsection*{Data and code availability}

\begin{itemize}
    \item All time-series scRNA-seq datasets used in this study have been previously reported and are publicly available. The dataset of zebrafish embryogenesis \citep{zebrafishData} was downloaded from \href{https://singlecell.broadinstitute.org/single_cell/study/SCP162/single-cell-reconstruction-of-developmental-trajectories-during-zebrafish-embryogenesis}{https:// singlecell.broadinstitute.org/single\_cell/study/SCP162}.  The dataset of reprogramming of mouse embryonic fibroblasts to induced pluripotent stem cells \citep{wotData} was downloaded from \href{https://broadinstitute.github.io/wot/tutorial/}{https:// broadinstitute.github.io/wot/tutorial/}. The dataset of pancreatic $\beta$-cell differentiation \citep{veresData} was downloaded in the NCBI under accession number \href{https://www.ncbi.nlm.nih.gov/geo/query/acc.cgi?acc=GSE114412}{GSE114412}.
    \item All original code is available at \href{https://github.com/QiJiang-QJ/scIMF}{https://github.com/QiJiang-QJ/scIMF}.
    \item Any additional information required to reanalyze the data reported in this paper is available from the lead contact upon request.    
\end{itemize}

\section*{ACKNOWLEDGMENTS}

This work was funded by the National Key Research and Development Program of China via[grant 2022YFA1004801.

\section*{AUTHOR CONTRIBUTIONS}

Conceptualization, Q.J. and L.W.; 
methodology, Q.J., L.Z., and L.W.; 
investigation, Q.J., L.Z., Lq.L., and L.W.; 
writing-–original draft, Q.J. and L.W.; 
writing-–review \& editing, Q.J., Lq.L., and L.W.;  
funding acquisition, L.W.;
resources, L.W.;
supervision, L.W.

\section*{DECLARATION OF INTERESTS}


The authors declare no competing interests.

\newpage

\bibliography{references}

\bigskip

\newpage

\section*{STAR METHODS}

\subsection*{Data Preprocess}
We benchmark scIMF against four state-of-the-art methods, PRESCIENT, MIOFlow, PI-SDE, and scNODE, using three time-series scRNA-seq datasets: zebrafish embryogenesis \citep{zebrafishData} (ZB data),  reprogramming of mouse embryonic fibroblasts to induced pluripotent stem cells \citep{wotData} (MEF data) and pancreatic $\beta$-cell differentiation \citep{veresData} (Panc data).

ZB data consists of 38,731 cells sampled across 12 tightly spaced developmental stages, spanning from zygotic genome activation to early somitogenesis. 
MEF data captures the transformation of mouse embryonic fibroblasts (MEFs) into induced pluripotent stem cells (iPSCs), with 236,285 cells profiled across 39 time points over 18 days.
Panc data provides a detailed view of stage 5 pancreatic $\beta$-cell differentiation and comprises a comprehensive transcriptomic profile of 51,274 cells classified into 12 unique cell types, spanning from Day 0 to Day 7.
We followed the data-preprocessing procedures as in scNODE, and downsampled MEF data to 10\% of their original sizes and performed temporal coarse-graining by grouping the 39 original time points into 19 time points, where each interval corresponds to a single day over the 18-day span.

For all tasks, we first identified the top 2,000 highly variable genes (HVGs) based on the training set and reduced data at all time points to 2,000 dimensions. We then applied UMI normalization (library size normalization to $10^4$) followed by log transformation. Next, we trained principal component analysis (PCA) using the training set and projected all time points into a 50-dimensional latent space using the learned principal components. Since scNODE operates in the HVG space, we first trained the model in the 2,000-dimensional space and then applied the same PCA transformation to project its 2,000-dimensional predictions into 50 dimensions for comparison with other methods.

\subsection*{Model multicellular dynamics as interacting diffusion processes}
Suppose the time series single-cell samples are collected at $(T+1)$ time points given by
\begin{equation}
	 \left( t_0, \mathbf{X}^{0}  \right), \left( t_1, \mathbf{X}^{1}  \right), \cdots,\left( t_T, \mathbf{X}^{T} \right) ,
\end{equation}
where $\mathbf{X}^{l} = \{\mathbf{x}^{l}_i\}_{i=1}^{N^l} \in \mathbb{R}^{N^l \times d}$ is a set of cells at a $d$-dimensional space either at the original gene expression space or the low-dimensional space from dimension reduction at time $t_l \;(0\le l\le T)$.  At each observed times,  cells are assumed to be drawn from a probability distribution in $d$-dimensional gene expression space, denoted by $\{\hat{\rho}_{t_i} \in \mathcal{P}(\mathbb{R}^d): i=0,1,\cdots,T\}$.

To model the collective behavior of the complex multicellular system, scIMF applies an interacting nonlinear diffusion process $\{\mathbf{x}_t \sim \rho_t : t_0 \le t \le t_T\}$ described by the MV-SDE:
\begin{equation}\label{eq:MVE}
\begin{aligned}
\mathrm{d} \mathbf{x}_t & =\mathbf{f}\left(\mathbf{x}_t, \rho_t\right) \mathrm{d} t+\sigma \mathrm{d} W_t, \quad t \in[t_0, t_T], \quad x_{t_0}=\rho_{t_0}, \\
\rho_t & =\text{Law}(\mathbf{x}_t), \quad \forall t_0 \le t \le t_T,
\end{aligned}
\end{equation}
where $\mathbf{f}\left(\mathbf{x}_t, \rho_t\right) $ is the interacting mean-field drift term,  $\{W_t\}_{t_0 \le t \le t_T}$ is a standard Brownian motion, $\sigma \ge 0$ is the strength of diffusion term, $\text{Law}(\mathbf{x}_t)$ represents the probability distribution of the random variable $\mathbf{x}_t$.

Mathematically, the drift term $\mathbf{f}$ for a particular cell is not only related to its own gene expression. Rather, the distribution of cells within the same time point is used as input to account for the influence of cell-cell interactions on cell fate decisions. Furthermore, the nonlinear dependence on $\rho_t$ is nonlocal, as $\mathbf{f}$ is a function not of the value $\rho(\mathbf{x}, t)$ but of the distribution $\left(\rho(\mathbf{x},t)\right)_{\mathbf{x}\in \mathbb{R}^d}$ of cells at time point $t$. 
Note that when the drift coefficient $\mathbf{f}$ is independent of $\rho_t$, Eq.~\ref{eq:MVE} reduces to describe a standard (It\^o-type) diffusion process. 
Thus, the MV-SDE generalizes classical SDEs by incorporating distribution-dependent dynamics, allowing for the modeling of cell-cell interactions.

We follow  \citeauthor{mishura2020existence}\citep{mishura2020existence} and decompose $\mathbf{f}$ into a non-interacting component and an interacting component, with the dependence on $\rho_t$ expressed as an expectation:
\begin{equation}\label{eq:f}
\begin{aligned}
	\mathbf{f}(\mathbf{x}_t,\rho_t) &= \mathbf{f}_{\text{intra}}(\mathbf{x}_t) + \mathbf{f}_{\text{inter}}(\mathbf{x}_t,\rho_t),\\
	&= \mathbf{f}_{\text{intra}}(\mathbf{x}_t) + \int_{\mathbf{y}_t}\varphi\left(\mathbf{x}_t, \mathbf{y}_t\right)\rho_t(\mathrm{d}\mathbf{y}_t).
\end{aligned}
\end{equation}
Here, the first term $\mathbf{f}_{\text{intra}}:\mathbb{R}^{d} \rightarrow \mathbb{R}^d$ represents the influence of each cell's individual state on the their dynamics. The second term is also called ``true McKean-Vlasov case'' \citep{mishura2020existence}, representing the influence of mean-field interactions. More specifically, $\varphi: \mathbb{R}^d \times \mathbb{R}^d \rightarrow \mathbb{R}^d$ quantifies the drift impact of an cell in state $\mathbf{y}_t$ on another cell in state $\mathbf{x}_t$. Then, the total impact on the cell in state $\mathbf{x}_t$, i.e., $\mathbf{f}_{\text{inter}}(\mathbf{x}_t,\rho_t)$, is integrating over the whole state space $\mathbb{R}^d$ weighted by the probability value. While this choice of dynamics is not the most general possible, it is sufficiently broad for many applications of interest \citep{sharrock2021parameter}.

\subsection*{Encode the interacting drift term of MV-SDE with Transformer}

Despite the theoretical richness of the MV-SDE, directly solving its full form poses significant computational challenges due to the high dimensionality of real data and complexity of the interaction terms. 
Therefore, we use neural networks to parameterize the specific form of the coefficient of MV-SDE (Eq.~\ref{eq:f}).
For $\mathbf{f}_{\text{intra}}$, we can simply utilize a two-layer fully connected neural network as the underlying architecture. However, for $\mathbf{f}_{\text{inter}}(\mathbf{x}_t,\rho_t)$, conventional neural networks, which process each cell's gene expression independently, fails to consider interactions between cells.
Here, we utilize the self-attention mechanism of Transformer, which allows for exploring the quadratic  relationships between tokens (e.g., cells in this study).
By regarding gene expressions of each cell as tokens, we implement a cell-wise attention mechanism to approximate the $\mathbf{f}_{\text{inter}}(\mathbf{x}_t,\rho_t)$.

More precisely, given a set of cells at time $t$, $\mathbf{X}^{t} = \{\mathbf{x}^{t}_i\}_{i=1}^{N} \in \mathbb{R}^{N \times d}$, we conduct the cell-wise attention mechanism through:
\begin{equation}\label{eq:self-attention-1}
\begin{aligned}
\mathbf{Y}^t&=\operatorname{Attention}\left(\mathbf{Q}^t, \mathbf{K}^t, \mathbf{V}^t\right),\\
&=\mathbf{A}^t\mathbf{V}^t,\\
&=\operatorname{softmax}\left(\frac{\mathbf{Q}^t(\mathbf{K}^t)^{\text{T}}}{\sqrt{d_k}}\right) \mathbf{V}^t,
\end{aligned}
\end{equation}
where
\begin{equation}\label{eq:self-attention-2}
\mathbf{Q}^t=\mathbf{X}^t\cdot \mathbf{W}^Q,\quad \mathbf{K}^t=\mathbf{X}^t\cdot \mathbf{W}^K, \quad \mathbf{V}^t=\mathbf{X}^t\cdot \mathbf{W}^V.
\end{equation}
Here, the parameter matrices $\mathbf{W}^Q \in \mathbb{R}^{d \times d_k}, \mathbf{W}^K \in \mathbb{R}^{d \times d_k}, \mathbf{W}^V \in \mathbb{R}^{d \times d}$ are learned during training, $\operatorname{softmax}\left(\mathbf{z}_i\right)=\frac{\exp \left(\mathbf{z}_i\right)}{\sum_j\exp\left( \mathbf{z}_j\right)}$. Notably, $\mathbf{W}^Q$ and $\mathbf{W}^K$ are trained independently without constraints enforcing their equality. As a result, the estimated attention scores can be asymmetric, making them well-suited for capturing the influence of source cells on target cell velocities. This further distinguishes our model from kernel-based methods \citep{DIISCO, NeuralMVP}, where the interaction matrix restricts to symmetric, implying directionless cell-cell interactions.

To decouple $\mathbf{f}_{\text{inter}}$ from $\mathbf{f}_{\text{non\_inter}}$, we ensure that $\mathbf{f}_{\text{inter}}$ only considers the interactions between cells, excluding self-influence. In the practical implementation, we achieve this by masking the diagonal elements of the attention matrix, ensuring that the attention scores for the diagonal elements  (i.e., the attention scores for each cell with respect to itself) are zeroed out.

In addition, interactions between cells can be complex and involve a variety of types (pathways). This can be represented by the multi-head mechanism in Transformer. Specifically, suppose there are $H$ interactions among the multicellular system, the multi-head self-attention is calculated as follows:
\begin{equation}\label{eq:multihead-attention1}
\mathbf{Y}^t =\text {Concat}(\textbf {head}^t_1, \ldots, \textbf {head}^t_H) \mathbf{W}^O,
\end{equation}
where
\begin{equation}\label{eq:multihead-attention2}
\begin{aligned}
\textbf{head}^t_h&=\operatorname{Attention}\left(\mathbf{Q}^t_h, \mathbf{K}^t_h, \mathbf{V}^t_h\right),\\
\mathbf{Q}^t_h=\mathbf{X}^t\cdot \mathbf{W}^Q_h,\quad & \mathbf{K}^t_h=\mathbf{X}^t\cdot \mathbf{W}^K_h, \quad \mathbf{V}^t_h=\mathbf{X}^t\cdot \mathbf{W}^V_h,\\
& h=1,2,\cdots,H.
\end{aligned}
\end{equation}
Here,  $\mathbf{W}^O \in \mathbb{R}^{H d \times d}$ is a linear transformation used to integrate the outputs of all attention heads. Each attention head contains its own weights $\{\mathbf{W}_h^Q \in \mathbb{R}^{d \times d_k}, \mathbf{W}_h^K \in \mathbb{R}^{d \times d_k}, \mathbf{W}_h^V \in \mathbb{R}^{d \times d} \}$, which can be learned from data. The outputs of all parallel attention heads are concatenated into a single matrix and transformed by the linear matrix $\mathbf{W}^O$ to give the output matrix $\mathbf{Y}^t$. 

The resulting matrix $\mathbf{Y}^t$ is then passed through a subsequent Feed-Forward Network (FFN), which consists of two linear transformations with a nonlinear activation function applied in between.
The final outputs approximate $\mathbf{f}_{\text{inter}}(\mathbf{x}_t,\rho_t)$, considering cell-cell interactions.

\subsection*{Solving the interacting diffusion process under the framework of dynamic OT}

Denote the parametrized neural network for $\mathbf{f}(\mathbf{x}_t,\rho_t)$ as $\mathbf{f}_{\theta}$, Eq.~\ref{eq:MVE} is further solved by optimizing the dynamic OT cost:
\begin{equation}\label{eq:dynamic-OT}
\begin{aligned}
 \inf_{\theta} \;\; & \mathbb{E}\left\{\int_{t_0}^{t_T}\|\mathbf{f}_{\theta}\left(\mathbf{x}_t, \rho_t\right)\|^2 \mathrm{d} t \right\},\\
\text{s.t. }\quad  \mathrm{d} \mathbf{x}_t & =\mathbf{f}_{\theta}\left(\mathbf{x}_t,\rho_t\right) \mathrm{d} t+\sigma \mathrm{d} W_t, \quad t \in[t_0, t_T], \\
\rho_{t} &= \text{Law}(\mathbf{x}_t),\\
 \mathbf{x}_{t_0} &\sim \hat{\rho}_{t_0},\\
 \rho_{t_l} &= \hat{\rho}_{t_l},\; l=1,2,\cdots,T.
\end{aligned}
\end{equation}
This involves ensuring that the modeled diffusion process aligns well with the observed data distribution while adhering to the principle of least action, which governs the natural evolution of biological systems \citep{PISDE}.

The marginal distribution constraints at observed time points can be overly restrictive, making direct optimization challenging. To address this,  these constraints are often relaxed by introducing a penalty term into the objective function. Specifically, a Wasserstein loss is added to quantify the discrepancy between the predicted and true distributions, allowing for a more flexible alignment. This relaxed formulation can be written as: 
\begin{equation}\label{eq:loss-Methods}
\begin{aligned}
 \inf_{\theta} \;\; & \mathbb{E}\left\{\int_{t_0}^{t_T}\|\mathbf{f}_{\theta}\left(\mathbf{x}_t, \rho_t\right)\|^2 \mathrm{d} t \right\} + \lambda \sum_{l=1}^T \mathrm{W}_2 \left( \hat{\rho}_{t_l}, \rho_{t_l}\right)^2\\
\text{s.t. }\quad  \mathrm{d} \mathbf{x}_t & =\mathbf{f}_{\theta}\left(\mathbf{x}_t, \rho_t\right) \mathrm{d} t+\sigma \mathrm{d} W_t, \quad t \in[t_0, t_T], \\
\rho_{t} &= \text{Law}(\mathbf{x}_t),\\
 \mathbf{x}_{t_0} &\sim \hat{\rho}_{t_0}.
 \end{aligned}
\end{equation}
where $\lambda \ge 0$ is a hyperparameter to balance the importance of satisfying the principle of least action and minimizing distributional differences at observed time points, $\mathrm{W}_2(\mu,\nu)$ represents the Wasserstein distance between distribution $\mu$ and distribution $\nu$. 

Finally, we formulate the entire problem as a MV-SDE-constrained optimization problem and solve it efficiently using the NeuralSDE framework, where a transformer encoder network approximates the interacting mean-field drift term of the MV-SDE.
Optimization of $\mathbf{f}_{\theta}$ is conducted using the Adam optimizer, with a batch size of 512. For the architecture of $\mathbf{f}_{\text{intra}}$, we utilize a fully connected 2-layer, 256-unit model employing leakyrelu as the activation function. As for $\mathbf{f}_{\text{inter}}$, we employ a single-layer transformer encoder with 2 attention heads and a dropout rate of 0.1 for regularization. The diffusion coefficient is prescribed as a constant, set at 0.1. During training, the model is configured with a learning rate of 0.001, and gradient clipping was applied with a maximum norm of 0.1. The Wasserstein distance is computed using the Sinkhorn algorithm \citep{cuturi2013sinkhorn}, with a scaling of 0.7 and a blur of 0.1.

\end{document}